# Home Heating Systems Design
# Using PHP and MySQL Databases

Assoc.Prof. Tiberiu Marius Karnyanszky, Ph.D., Dipl.Eng., Dipl.Ec.
"Tibiscus" University of Timişoara, România

**ABSTRACT:** Modificările climatice din ce în ce mai evidente din zilel noastre au dus, printre altele, la o avalanşă de cereri de instalare în locuinţa de domiciliu sau cea de vacanţă a unor dispozitive menite să mărească gradul de confort, printre care instalaţiile de încălzire proprie. Alegerea unuia sau altora dintre modelele existente pe piaţă se face după varii criterii, atât obiective cât şi subiective. Pentru furnizorul de aparatură de încălzire, această alegere trebuie să ţină cont de câteva elemente specifice cum sunt localizarea clădirii, orientarea sa, dimensiunile şi detaliile constructive. Lucrarea prezintă o aplicaţie informatică accesibilă pe Internet care dimensionează puterea instalaţiei de încălzire şi ghidează clientuk în alegerea unei centrale termice potrivit dorinţelor sale.

## 1. Generalities

Structures have different shapes and constructive characteristics, therefore the heating systems' design needs a computing of the heating necessary, expressing the quantity of thermal energy conceded by each room to the external medium.

This calculus of the heating necessities is made by two ways:
1-basic calculus, based on mathematical expressions, containing the heat flow conceded to the outside because of the temperature difference, the heating necessary of the cold air entered in the structure and the correspondent additions. It is an extremely rigorous and exactly calculus influenced by all the parameters which induces the heating necessary.

2-coefficent-based calculus, an approximate calculation based on the building type, the shape and geometrical dimensions, the isolation and climatic area of the structure.

This paper presents the use of a computer application based on a MySQL database, managed by PHP programs, allowing the selection of a heating device using coefficient-based calculus.

## 2. The mathematical model

For structures and similarly, the mathematical model of the heating necessary is:

$$Q = V * GN * (t_{mi} - t_e) \quad [W] \quad (1)$$

where:
- V -the interior volume of the structure, inside the envelope [$m^3$];
- GN -the global normalized coefficient of thermal isolation, determined by the floors number N and by the arithmetical ratio between building surface A and building volume V [$W/m^3*K$];
- $t_{mi}$ -the medium temperature of the inside air [C]
- $t_e$ -the medium temperature of the outside air [C]

**Table 1. The $G_N$ coefficient (fragment)**

| Levels (N) | Surface/Volume [$m^2/m^3$] | $G_N$ [$W/m^3K$] |
|---|---|---|
| 1 | 0,80 | 0,77 |
|   | 0,85 | 0,81 |
|   | 0,90 | 0,85 |
|   | 0,95 | 0,88 |
|   | 1,00 | 0,91 |
|   | 1,05 | 0,93 |
|   | ≥1,10 | 0,95 |
| 2 | 0,45 | 0,57 |
|   | 0,50 | 0,61 |
|   | 0,55 | 0,66 |
|   | 0,60 | 0,70 |
|   | 0,65 | 0,72 |
|   | 0,70 | 0,74 |
|   | ≥0,75 | 0,75 |

The structure volume V depends on the building dimensions (length * width * height) considering only the interior volume; the GN coefficient is listed (see Table 1); the outside temperature (see Table 2) and the inside temperature (see Table 3) are also listed and depends on the destination of the structure respectively the area where the building is placed.

**Table 2. Medium outside temperatures in Romania (fragment)**

| City | Temperature [C] | City | Temperature [C] |
|---|---|---|---|
| Alba Iulia | -18 | Lugoj | -12 |
| Arad | -16 | Oradea | -15 |
| Beiuş | -18 | Petroşani | -18 |
| Braşov | -21 | Reşiţa | -12 |
| Bucureşti | -15 | Sibiu | -18 |
| Cluj Napoca | -18 | Timişoara | -15 |
| Deva | -15 | Târgu Jiu | -15 |
| Hunedoara | -15 | Drobeta Tr Severin | -12 |

**Table 3. Medium inside temperatures – structures**

| Destination | Temperature [C] |
|---|---|
| Rooms and lobbies | 20 |
| Vestibules | 16 |
| Bathrooms | 22 |
| Kitchens | 16 |
| Toilets | 18 |
| Stairs | 10 |
| Entrances | 10 |
| Laundries and ironings | 15 |
| Drying rooms | 25 |
| Garages | 10 |

### 3. Heating devices

The structure heating and the warm water furnishing are made using two major ways: centralized regime, when the estimation of the heating necessary is made by the designer, respectively in personal regime, when the estimation is made by the owner which can use the following devices:
- condensing boilers – wall-hung or floor standing boilers, using methane gas, liquidated petroleum gas or gas oil;

- forced draught steel boilers using methane gas, gas oil or combined methane/oil gas;
- pig iron boilers using atmospheric burner with methane gas;
- solid fuel boilers using wood or sawdust;
- boilers and domestic hot water pumps;
- commercial boilers.

By choosing these boilers, user save energy, contribute to preserving the environment and increase their own comfort.

- **Save energy -** Economies by using only what is needed when it is needed.
- **Protect the environment -** Previously, domestic heating represented 32% of greenhouse gas emissions. To reduce this percentage, the aim is to adapt heating systems using ecological techniques and to increase the use of renewable energy sources.
- **Increase users comfort -** Live in harmony with nature thanks to technology which monitors the quality of water, respects the environment and reduces waste.

## 4. Computer application

This application is based on a MySQL database, running on the server, containing data about the thermo stations of one or all producers, data referring to main important characteristics (for example, see Table 4).
-station type, steam type and burner type
-installation possibilities
-power and consumption
-type of combustible
-environmental costs

The use of this application starts with the input of the parameters needed by the heat calculus using the coefficient-based method as (Figure 1):
-the structure's area, to obtain the medium exterior temperature;
-the aim of the structure, to obtain the medium interior temperature;
-the number of floors and the arithmetical ratio between building surface A and building volume V, to obtain the GN coefficient;
-the surface and the height of the structure, to obtain the structure's volume.

## Table 4. Uno-3 heating device characteristics

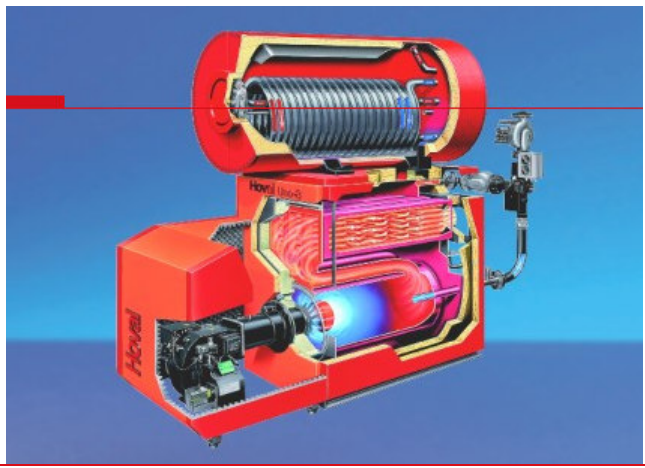

### Uno-3
**Futuristic heating boiler technology**
Choosing Hoval Uno means making a sound investment in futuristic heating boiler technology. The Uno type series meets the most stringent statutory requirements with regard to environmentally-friendly, low-emission economic operation. The finely-graded model range from 50 to 360 Kilowatt enables the installation to be exactly tailored to the purpose. Hoval has great experience precisely in this medium capacity segment and enjoys worldwide recognition for high-quality heating boilers.
**Kelps keep down operating and heating costs:** Standard efficiency of 95-96%.
**Promises lower environmental costs:** Minimum emission values under all operating conditions.
**Guarantees low installation costs:** Slender construction and a high degree of flexibility when installing.
**Advantages at a glance**
**Saves time on installation due to high flexibility:** Its size means that it can normally be installed without difficulty, otherwise on-site position welding is offered (naturally with a full works guarantee of functional safety).
**Thermolytic heating surface:** The self-cleaning effect prevents soot deposits and the high efficiency is maintained almost without reduction over a long operating period.
**Triple-draught flue gas system:** Based on the proven design principle of the Hoval Mini-3. Guarantees clean combustion.
**Operation:** The Hoval TopTronic T increases comfort and reduces energy costs due to its optimum control system.
**Compact central heating systems can also be supplied.**
**Facts**
**Environment:** Lowest emissions in all operating conditions.
**Energy consumption:** Standard efficiency of 95-96%.
**Performance levels:** Thirteen performance levels of 35 - 360 kW

**Figure 1. Heating necessities calculus**

After the data input, the application determines the power of the heating device, in kW and MCal, as presented in Figure 2.

**Figure 2. The calculated value for the heating power**

Next, the MySQL database furnishes the available devices to support the calculated power of the heating steam, presenting the main characteristics (Figure 3).

The user can now input other characteristics of the heating device to reduce the number of possibilities, as steam, burner and combustible type (Figure 4) so the available list is shorter and the options of the user are reduced to the most valuable ones.

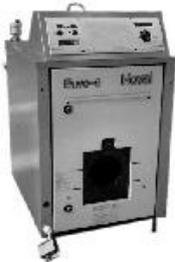
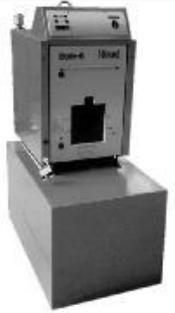
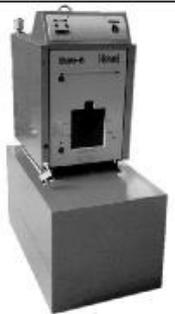

**Figure 3. The selected devices from the database**

**Conclusions**

This paper presents a computer application, accessible to all users that have an Internet connection, allowing selecting a device able to heat, at a desired

127

temperature, a structure placed into a selected location and having some structural characteristics.

**Figure 4. The particularization of the device's selection**

The paper can be followed by the calculus of the power to heat the domestic warm water, and so this application is useful to size a device used by the owner.
The usage of databases and of a database management system, accessible by Internet, is very utile for design companies, engineering companies, to the future owners but also for the heating producers' representatives.

128